# Do Users Want Platform Moderation or Individual Control? Examining the Role of Third-Person Effects and Free Speech Support in Shaping Moderation Preferences

Shagun Jhaver, Amy Zhang

This study examines social media users' preferences for the use of platform-wide moderation in comparison to user-controlled, personalized moderation tools to regulate three categories of norm-violating content – hate speech, sexually explicit content, and violent content. Via a nationally representative survey of 984 US adults, we explore the influence of third-person effects and support for freedom of expression on this choice. We find that perceived negative effects on others negatively predict while free speech support positively predicts a preference for having personal moderation settings over platform-directed moderation for regulating each speech category. Our findings show that platform governance initiatives need to account for both actual and perceived media effects of norm-violating speech categories to increase user satisfaction. Our analysis suggests that users do not view personal moderation tools as an infringement on others' free speech but as a means to assert greater agency over their social media feeds.

# Introduction

With the emergence of social media sites and their widespread use for interpersonal communication, companies like Facebook, Twitter, and YouTube have become the new governors of digital expression. Concurrently, individuals using these sites can also give their input into governance in several ways. For example, they can flag posts that violate community policy, downvote inappropriate posts, serve as volunteer moderators, engage in counter-speech, or configure moderation settings to remove certain posts automatically. We are therefore moving towards a "pluralist model of speech regulation (Balkin, 2017)," in which speech must be regulated in a multi-stakeholder fashion – *legislative entities* enforce online speech laws, *platform operators* configure governance regimes of acceptable content, and *users themselves* intervene against content perceived as problematic.

This move toward a pluralist model is occurring in the context of recent controversies over platforms' moderation decisions (Stecklow, 2018; Gillespie, 2010; Massanari, 2017) and growing media, policymaker, and public calls to regulate their content better (Williams et al., 2021). In response, platforms have begun investing more resources into improving how inappropriate posts are detected and removed from their sites. We focus in this article on platforms' offering of *personal moderation tools* that let end-users configure content moderation of the posts they see to align with their content preferences. We are primarily concerned with tools offered by platforms such as Instagram and Twitter that let users specify their sensitivity to specific topical categories, such as sexually explicit content and hate speech (see Figure 1). We note a distinction between who identifies a given content as belonging to a topical category and who decides whether to remove content from that category. While the former is a critical, contentious issue, our focus here is on the latter

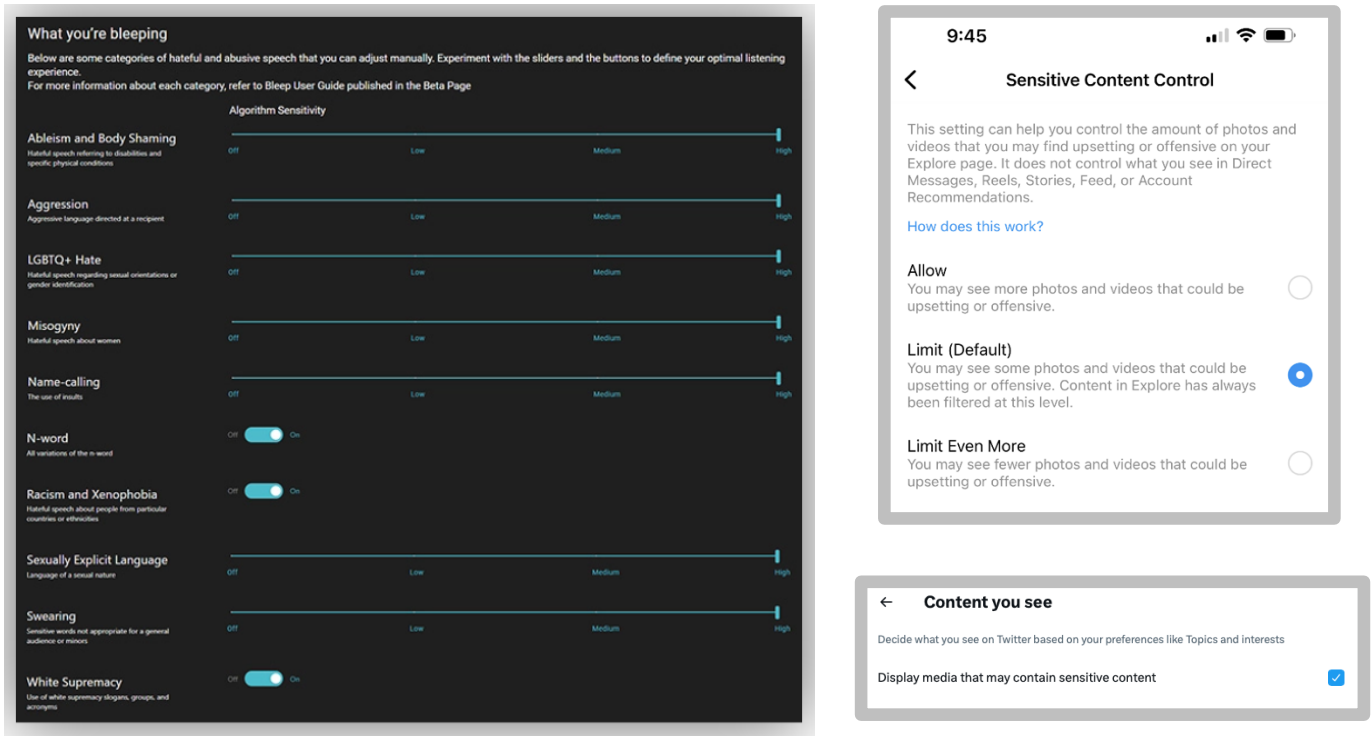

*Figure 1: Examples of personal moderation tools on Intel Bleep (left), Instagram (top right), and Twitter (bottom right).*

concern.

Configuring such tools lets users align the moderation system with their tastes and thresholds. From the perspective of platforms, a tactical consequence of offering personal moderation tools is ceding to users the responsibility of making hard moderation decisions, the concomitant onus of making mistakes with them, and the cognitive labor of making the 'correct' configurations. Therefore, it is vital that researchers understand how users consider the choice between platform versus personal moderation.

Our research responds to calls by governance scholars to conduct more survey-based research to understand users' perspectives on moderation interventions by different regulatory actors (Einwiller and Kim, 2020; Riedl et al., 2021a). To date, we have scant knowledge of how end-users perceive being given self-regulating authority through personal moderation tools. We do not know the situations in which users would prefer to have a choice in shaping moderation versus when they would instead prefer that platforms manage it for all users – and the different factors that shape these preferences.

Informed by the third-person effects (TPE) hypothesis, we fill this gap by examining users' preferences in the context of three norm-violating speech categories previously studied in the literature: (1) hate speech, (2) sexually explicit content, and (3) violent content. Prior research has shown that *perceptions* of the effects of media messages *on others* predict content regulation attitudes (Riedl et al., 2021b; Rojas et al., 1996). We examine the role that TPE (Davison, 1983) plays in shaping user attitudes about deploying platform-enacted versus personal moderation tools.

We also connect our findings to the scholarship on understanding public attitudes toward freedom of expression and its consequences. Many Americans highly value free speech, as evidenced by the First Amendment. However, the introduction of personal moderation tools complicates upholding free speech principles. Users may use these tools to avoid specific content categories while others continue to see the same content, thereby preventing infringement of online expressions. On the other hand, personal moderation tools could be framed as a way for people to avoid viewpoints they dislike. While one could always ignore any content, personal moderation tools make removing broad swathes of content easy. Beyond eschewing the use of personal moderation tools themselves, free speech advocates may prefer that others did not have the option to use them. Therefore, we analyze how users' support for freedom of expression shapes their notion of different moderation approaches.

Understanding public views on platform- and user-enacted interventions can stimulate debates about the roles and strategies of various regulatory actors (Seering et al., 2020; Seering et al., 2019). It can also address calls for evidence-based policymaking (Livingstone, 2013; Puppis and Van den Bulck, 2019) by clarifying how the public understands moderation practices and identifying gaps between policy and public demands (Naab et al., 2021; Riedl et al., 2021a; Paek et al., 2008). Given the rapid introduction of new moderation strategies by the platforms, especially personal moderation tools, independent academic assessments of users' attitudes on their deployment are vital. Platforms can also benefit from such research by understanding end-users' acceptance or rejection of various regulatory practices and the factors that shape those perspectives toward building better tools. Further, examining the public perception of free speech within the context of online activity may also shape attempts to protect free speech in the long run.

## Literature Review and Hypotheses

### The Perceptual Component of TPE

In 1983, Davison first argued that individuals perceive media's impact on the attitudes and behaviors of others to be greater than it is on themselves (Davison, 1983). Since then, many studies have shown this discrepancy to be consistent across a range of contexts (Sun et al., 2008), such as political ads (Golan et al., 2008; Paek et al., 2005), news stories (Price and Tewksbury, 1996), and social media use (Schweisberger et al., 2014). This hypothesis, which Davison called the *third-person effect (TPE)* (Davison, 1983), has become a widely applied perspective to explain public opinion on media censorship (Gunther, 2006; Guo and Johnson, 2020; Hoffner et al., 1999; McLeod et al., 1997). The TPE hypothesis has two major components – perceptual and behavioral (Gunther, 1991). The perceptual component states that presumed media effects on others (*PME3*) tend to be greater than perceived media effects on self (*PME1*). In the context of social media messages' influence, the perceptual component predicts that individuals will consider others to be more negatively influenced by each category of norm-violating speech on social media than themselves. We, therefore, raise the following hypothesis:

*H1:* For each norm-violating speech category, participants will perceive a greater effect of that speech on others than on themselves.

## The Behavioral Component of TPE

The behavioral component of TPE posits that when individuals perceive the greater impact of media messages on others than on themselves, they will take remedial actions to mitigate the perceived harmful effects (Guo and Johnson, 2020). Davison described the phenomenon of censorship as one of the most interesting behavioral consequences of third-person perception (Davison, 1983). Prior research on TPE consequences shows that it leads to censorship support (Rojas et al., 1996; Gunther, 2006; Golan and Banning, 2008), and the effects are particularly salient when persuasive attempts may include socially undesirable effects (Lim, 2017). In studying TPE effects on censorship attitudes, some researchers have examined the consequences of the other-self perceptual disparity in media effects (*DME* = PME3 – PME1). Others have focused on the perceived media impact on others (PME3) (Riedl et al., 2021b). Examining past research data on TPE consequences, Chung and Moon concluded that the media's presumed effect on others (PME3) is a stronger predictor of censorship attitudes than the other-self disparity in the perceived media effects (Chung and Moon, 2016). In this research, we choose PME3 as our primary predictor variable and consider the following hypothesis:

*H2:* For each speech category, the perceived effects of that speech on others will be positively related to support for the platform's banning of that category.

Researchers of TPE have long been curious about the potential behaviors that could result from the perceived media impact on others. In addition to censorship support, prior studies have investigated user behavioral outcomes, such as engaging in political action (Feldman et al., 2017; Rojas, 2010; Wei et al., 2011), disseminating opposing information (Bernhard and Dohle, 2015; Golan and Lim, 2016), and exposing apparent biases (Barnidge and Rojas, 2014; Lim and Golan, 2011). In the context of online content moderation, Jang and Kim found that platform users with a greater level of third-person perception were more likely to support media literacy interventions to address fake news (Jang and Kim, 2018). We add to previous efforts to surface the different types of behavioral consequences of TPE by examining its impact on users' support for having personal moderation settings to moderate norm-violating content. Since such configurations are also a form of regulation, i.e., end-user regulation of content, prior TPE research suggests that PME3 would predict support for them (Gunther, 2006; Riedl et al., 2021b). This research, therefore, considers the following hypothesis:

*H3:* For each speech category, the perceived effects of that speech on others will be positively related to support for having personal moderation tools to regulate the speech of that category.

Platform-enacted moderation and personal configurations to self-moderate content are different ways to regulate norm-violating speech. However, while platform-enacted moderation censors content platform-wide, personal moderation tools allow users to adjust whether and how much norm-violating speech they are willing to encounter personally. Prior literature generally predicts that TPE would lead to support for censorship attitudes (Gunther, 2006; Chung and Moon, 2016). However, it does not guide how people will react to a choice between letting platforms handle a specific content category and allowing users to specify their moderation preferences for that category. In this research, we consider the following research question:

*RQ1:* For each speech category, how do its perceived effects on others relate to support for the platform's banning of that category versus support for having personal moderation tools regulate it?

## Support for Freedom of Speech

Discussions about the benefits of moderation measures are always intertwined with the issue of *freedom of speech*. In the United States, the Constitution protects the right to free expression as a fundamental human right. However, platforms are private parties. Section 230 of the Communications Decency Act of 1996 provides them the legislative freedom to police their users as they see fit while not being held accountable for errors or oversights (Gillespie, 2017; Medeiros, 2017). Experts have shown that despite Americans' support for freedom of expression generally, their tolerance for hate speech is low (Gibson and Bingham, 1982; Sullivan et al., 1993; Yalof and Dautrich, 2002). Thus, people's acceptance of free speech in the abstract may not automatically imply their tolerance for opposing expressions (Naab, 2012). Examining the degree to which people's support for free speech affects their opinions about varied moderation strategies may help to better explain this discrepancy in the context of social media platforms.

Support for free speech and attitudes toward content moderation have been linked in previous studies. For instance, Naab et al. showed that people who commit to freedom of expression are less likely to support restrictive actions by Facebook moderators (Naab et al., 2021). Guo and Johnson showed that a lack of support for freedom of speech predicts support for government regulation of sexist hate speech (Guo and Johnson, 2020). However, they did not find that the former could predict supportive attitudes toward platform censorship. Jang and Kim suggested that support for free expression decreases support for regulating fake news despite the existence of third-person effects (Jang and Kim, 2018). Overall, this body of research indicates that participants' support for platform regulation will decline as free speech support increases. In this research, we consider the following hypothesis:

*H4:* For each speech category, participants' support for freedom of expression will be negatively related to their support for the platform's banning of that category.

The sparse literature on personal moderation tools offers no direct guidance on the relationship between support for free speech and support for having such tools – however, some prior work addresses related issues. For example, Naab et al. found no relationship between users' commitment to free speech and their intention to engage in corrective actions, such as rebuking the comment author or reporting the comment (Naab et al., 2021). It is unclear whether that finding would apply to our context since deploying personal moderation tools is not a corrective action in this sense. Its expected costs are also lower than corrective actions, e.g., users may set personal moderation configurations once instead of reporting every inappropriate post; personal moderation privately removes problematic comments and avoids confrontations with the comment authors whereas engaging in counter-speech may invite retaliation. A Riedl et al. survey showed that individuals' support for free speech does not increase their sense of obligation to intervene against problematic comments (Riedl et al., 2021a); however, the authors did not tell the survey takers their intervention options. Further, in our study context, users may not consider personal moderation tools an obligation.

On the one hand, support for free speech values should be expected to reduce support for most, if not all, restrictive actions. On the other hand, people may perceive personal moderation tools as simply offering them greater agency to shape what they see, not as an infringement on the free speech of others. In this way, personal moderation tools allow a distinction between one's freedom to speak and one's obligation to be heard by others. To clarify the direction of this relationship, this research poses the following question:

*RQ2:* Does participants' support for freedom of expression relate to their support for having personal moderation tools?

When faced with a choice between letting platforms handle the moderation of a certain content category and letting users specify their own moderation preferences, we expect the latter to be perceived as more free speech preserving. It accords users, not platforms, the freedom to decide whether and how much content to remove. We, therefore, raise the following hypothesis:

*H5:* For each speech category, participants' support for freedom of expression will be related to greater support for having personal moderation tools to moderate that category than their support for the platform's banning of that category.

## Methods

Our study was considered exempt from review by the Institute[i] IRB. We recruited participants through Lucid,[ii] a survey company that provides researchers access to nationally representative samples. Our inclusion criterion for survey participants was all adult internet users in the US. We paid participants through the Lucid system.

We designed our online survey questions to test previously noted hypotheses and answer our research questions. We adapted survey instruments from relevant prior literature to assess some measures, as described below. To increase survey validity, we sought feedback on an early survey questionnaire draft from students involved in social computing research at the authors' institutions. Nine students who were not involved with the project provided feedback on the wording of the questions and the survey flow, which we incorporated into the final survey design. We also piloted the survey with 27 participants through Lucid. During this pilot test, we included this open-ended question at three different points in the survey: "Do you have any feedback on any of the questions so far? For example, is any question unclear or ambiguous? Please list the question and describe your challenge with answering it." At the end of the survey, we also asked, "Overall, how can we improve this survey from the perspective of survey takers? Do you have any other thoughts or feedback for us? Please describe." Results from this survey pretest resulted in another round of iteration.

Our survey questionnaire contained three blocks with similar questions about hate speech, sexually explicit content, and violent posts. To counter the effects of the order of presentation on survey results, we counterbalanced the sequence in which the three question blocks were shown to participants. At the beginning of each block, we specified the norm-violating category to which the following questions related and defined that category using the following definitions:

- *Hate speech:* "Hate speech includes speech that is dehumanizing, stereotyping, or insulting, on the basis of identity markers such as race/ethnicity, gender, sexual orientation, religion, etc."

- *Violent content:* "Violent content includes threats to commit violence, glorifying violence or celebrating suffering, depictions of violence that are gratuitous or gory, and animal abuse."

- *Sexually explicit content:* "Sexually explicit content includes content showing sexual activity, offering or requesting sexual activity, female nipples (except breastfeeding, health, and acts of protest), nudity showing genitals, and sexually explicit language."

The preceding definitions were inspired by the language on the Facebook site when reporting hate speech, violence, and nudity posts. We administered the survey online using the survey software package Qualtrics. We launched the survey on 29 November 2022. Table 1 presents the demographic details of our final sample after data cleaning and compares them to the demographics of the adult US internet population (Ruggles et al., 2022).

*Table 1: Demographic Profile of the US Survey*

|  | Authors' study, US survey Nov 2022 (%) | American Community Survey, US sample 2021 (%) |
|---|---|---|
| *Age:* |  |  |
| 18-29 | 18.7 | 17.4 |
| 30-49 | 40.1 | 29.5 |
| 50-64 | 23.7 | 25.6 |
| 65+ | 17.4 | 27.3 |
| *Gender:* |  |  |
| Male | 48.2 | 48.6 |
| Female | 51.8 | 51.4 |

| Race/Ethnicity: | | |
|---|---|---|
| White | 73.5 | 68.3 |
| Black | 12.2 | 9.3 |
| Other | 14.3 | 22.4 |
| Hispanic: | | |
| Yes | 4.5 | 13.7 |
| Education: | | |
| High school or less | 31.4 | 33.5 |
| Some college | 24.4 | 33.3 |
| College+ | 43.3 | 33.1 |

## Measures

### Perceived Influence on Self and Others

For each norm-violating speech category, we asked participants to estimate the influence of that category on the self and others. In each case, we posed two questions adapted from (Rojas et al., 1996): "Seeing <speech category> posts on social media has a powerful influence on my attitudes." and "Seeing <speech category> posts on social media has a powerful influence on my behaviors." The response categories ranged on a 7-point Likert scale from 1 (strongly disagree) to 7 (strongly agree). These two items were averaged to create a measure of the perceived influence of each speech category on the self (Hate speech: $M=3.89$, $SD=1.91$, $\alpha=.84$; Violent speech: $M=3.78$, $SD=1.87$, $\alpha=.81$; Sexually explicit speech: $M=3.55$, $SD=1.89$, $\alpha=.87$). We asked another two questions that replaced only the word "my" with "other people's" and averaged the responses to create an index of the perceived influence of each speech category on others (Hate speech: $M=5.28$, $SD=1.46$, $\alpha=.92$; Violent speech: $M=5.19$, $SD=1.40$, $\alpha=.91$; Sexually explicit speech: $M=4.99$, $SD=1.46$, $\alpha=.93$).

### Support for Freedom of Speech

We used items developed by Guo and Johnson to measure support for freedom of speech (Guo and Johnson, 2020). Participants rated four statements on a Likert scale ranging from 1 (strongly disagree) to 7 (strongly agree): (1) "In general, I support the First Amendment." (2) "Freedom of expression is essential to democracy." (3) "Democracy works best when citizens communicate in an unregulated marketplace of ideas." (4) "Even extreme viewpoints deserve to be voiced in society." We included the text of First Amendment[iii] in the first question to clarify its meaning. We formed an index for free speech support using the means of these four items (M=5.43, SD=1.15, α=.80).

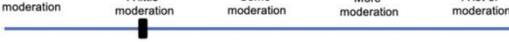

*Figure 2: Survey question asking participants to rate their support for platforms providing a personal moderation feature.*

## Dependent Variables Related to Moderation

*Support for platform-enacted moderation* of each speech category was operationalized by asking participants to rate the following statement: "I support social media platforms taking down any posts they consider to be <speech category> so that no users can see them." The responses to this statement ranged on a 7-point Likert scale from 1 (strongly disagree) to 7 (strongly agree).

To operationalize support for having personal moderation tools for each category, we showed participants an example of a personal moderation feature where every user can decide the extent to which they want to filter out the <speech category> (see Figure 2). We asked participants to rate their support for providing this kind of setting to all users on a Likert scale ranging from 1 (strongly disagree) to 7 (strongly agree).

In addition to these two measures, we also operationalized *choosing platform-enacted moderation vs personal moderation* for each speech category by asking respondents a binary question: "Given a choice between platform-wide moderation and a 'Choose your moderation settings' feature to handle <speech category> posts, which would you prefer to have?" The response categories included: (1) "*Platform-wide moderation:* Platforms should have the power to remove all posts they identify as <speech category> across the platform," or (2) "*'Choose your moderation settings' feature*: Each user should be allowed to configure the extent to which <speech category> posts should be removed for them." We note that the "Choose your moderation settings" feature enables a range of choices – from "no moderation" to "a lot of moderation" as shown in Figure 2. Thus, users who desired neither platform-wide nor personal moderation to remove posts for a speech category could select this feature and configure it at the "no moderation" level.

## Control Variables

Prior research has shown that socio-demographic variables are related to TPE, free speech, and attitude toward media regulation (Lambe, 2002; Gunther, 2006; Lee, 2009; Lo and Chang, 2006). Further, social media use has also been associated with individuals' perceptions of harmful content and their interventions against such (Naab et al., 2018; Kalch and Naab, 2017; Kenski et al., 2020; Ziegele et al., 2020). Therefore, we controlled for age, education, gender, race, political affiliation (1 = "strong Democrat", 7= "strong Republican"), and social media use of each respondent. We operationalized the frequency of social media use by following recommendations by Ernala et al. (Ernala et al., 2020) and prompting participants to respond to the question, "In the past week, on average, approximately how much time PER DAY have you spent actively using any social media?"

## Results

Our results show that 72.8%, 73.1%, and 66.2% of participants at least somewhat agreed that platforms should ban hate speech, violent content, and sexually explicit content, respectively. Further, 69.9%, 72.6%, and 72.1% of participants at least somewhat agreed that platforms should offer personal moderation tools to let end-users regulate hate speech, violent content, and sexually explicit content, respectively (see Figure 3).

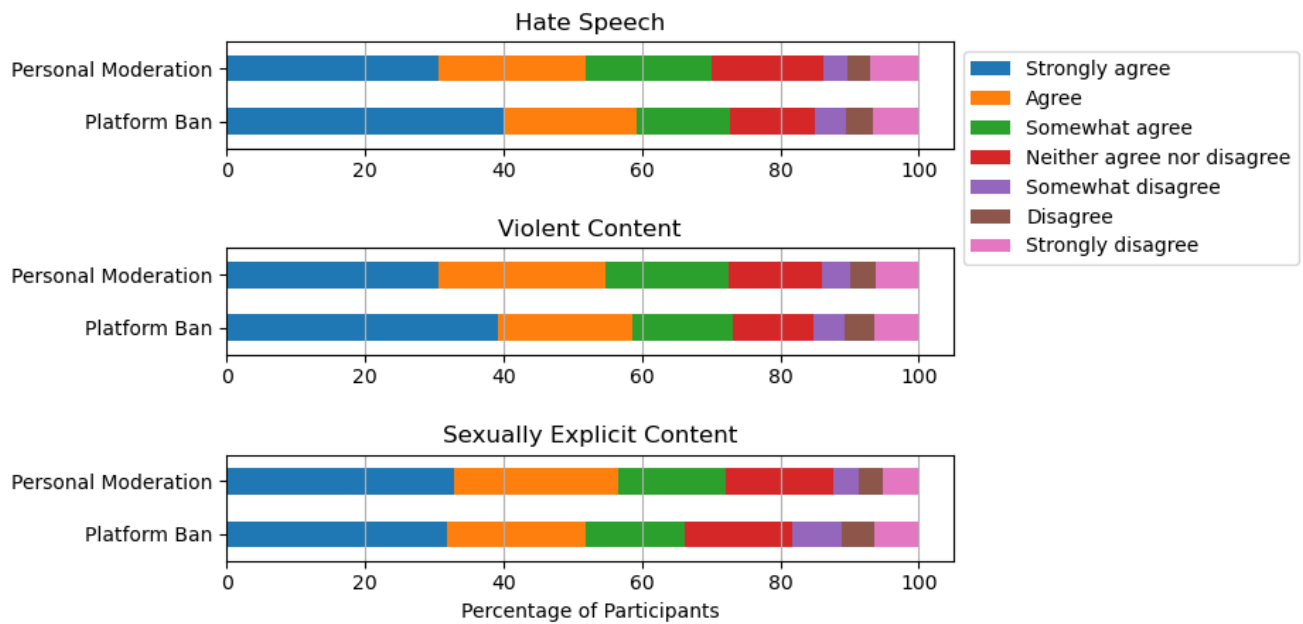

*Figure 3: Frequency of participants' responses to survey questions about support for platform-wide and personal moderation, measured in percentage.*

In line with research on the third-person effects, *H1* predicted that for each norm-violating content category, participants would perceive the effects of that category on others to be stronger than on themselves. We ran a paired *t*-test and found the perceived effects on others significantly stronger than on oneself for each category (see Table 2). Thus, our results support H1.

*Table 2: Mean, standard deviations, standard errors of participants' perceived effects of hate speech, violent content, and sexually explicit content on others and self, and t-test results comparing perceived effects on others and self for each speech category (N = 993). \*\*\* denotes p < .001*

|  |  | M | SD | SE | t | Cohen's d |
|---|---|---|---|---|---|---|
| Hate speech | Effects on others | 5.28 | 1.46 | .05 | 25.37*** | .805 |
|  | Effects on self | 3.89 | 1.91 | .06 |  |  |
| Violent content | Effects on others | 5.19 | 1.40 | .05 | 25.44*** | .807 |
|  | Effects on self | 3.77 | 1.87 | .06 |  |  |
| Sexually explicit content | Effects on others | 4.99 | 1.46 | .05 | 25.81*** | .819 |
|  | Effects on self | 3.55 | 1.88 | .06 |  |  |

## Support for Platform Bans

We computed hierarchical linear regression to test our hypotheses 2 and 4. For each norm-violating category, we created a model where the participant's support for platforms banning that category served as the dependent variable. In Step 1 of the three regression models, we included the control variables age, gender, education, race, political affiliation, and social media use. In Step 2, we introduced the independent variables PME3 (perceived effects on others) for that category and support for free speech (Table 3).

*Table 3: Hierarchical multiple regression analyses predicting support for platforms' banning of hate speech, violent content, and sexually explicit content (N = 983).*

| Independent Variable | Support for platform ban of hate speech (β) | Support for platform ban of violent content (β) | Support for platform ban of sexually explicit content (β) |
|---|---|---|---|
| Step 1 |  |  |  |
| Age | .119*** | .055 | .102** |
| Gender (Female) | .127*** | .167*** | .153*** |
| Race (White) | .018 | .013 | -.059 |
| Education[a] | -.008 | .039 | -.022 |
| Political affiliation[b] | -.156*** | -.09** | .019 |
| Social media use[c] | .022 | .013 | .019 |
| $R^2$ | .093*** | .065*** | .05*** |
| Step 2 | | | |
| Support for free speech | -.101*** | -.039 | -.006 |
| Perceived effects of hate speech on others | .476*** | - | - |
| Perceived effects of violent content on others | - | .397*** | - |
| Perceived effects of sexually explicit content on others | - | - | .391*** |
| $R^2$ change | .215 | .15*** | .149*** |
| Total $R^2$ | .307*** | .215*** | .199*** |

**$p < .01$, ***$p < .001$ ($t$ test for $\beta$, two-tailed; F test for $R^2$, two-tailed).

[a]0= Less than secondary education; 1= Secondary education or more.

[b]1= Strong Democrat, 7= Strong Republican.

[c]1= Less than 10 minutes per day, 6= More than 3 hours per day.

$\beta$ = Standardized beta from the full model (final beta controlling for all variables in the model).

For each norm-violating speech category, the regression models show significant influences of the participants' perceived effects of that category on others (PME3) on their support for a platform ban of that category (Model 1: hate speech – $\beta = .476$, $p < .001$; Model 2: violent content – $\beta = .397$, $p < .001$; Model 3: sexually explicit content – $\beta = .391$, $p < .001$), supporting H2.

Greater support for free speech significantly negatively influences support for platform bans of hate speech ($\beta = -.101$, $p < .001$). It does *not*, however, influence support for platform bans of violent content ($\beta$ = -.039, $p$ > .05) or sexually explicit content ($\beta$ = -.006, $p$ > .05). Thus, H4 is only partially supported.

## Support for Personal Moderation

We computed hierarchical linear regression to test our hypothesis 3 and answer RQ 2. For each norm-violating category, we created a model where the participant's support for having personal moderation tools to regulate that category served as the dependent variable. In Step 1 of the three regression models, we included the control variables age, gender, education, race, political affiliation, and social media use. In Step 2, we introduced the independent variables PME3 (perceived effects on others) for that category and support for free speech (Table 4).

*Table 4: Hierarchical multiple regression analyses predicting support for using personal moderation tools (PMT) to regulate hate speech, violent content, and sexually explicit content (N = 983).*

| Independent Variable | Support for PMT to regulate hate speech ($\beta$) | Support for PMT to regulate violent content ($\beta$) | Support for PMT to regulate sexually explicit content ($\beta$) |
| --- | --- | --- | --- |
| Step 1 | | | |
| Age | .014 | -.017 | -.010 |
| Gender (Female) | -.016 | .019 | -.016 |
| Race (White) | -.019 | -.007 | .08* |
| Education[a] | -.018 | .022 | .003 |
| Political affiliation[b] | -.019 | -.049 | -.074* |
| Social media use[c] | .051 | .057 | .086** |
| $R^2$ | .009 | .013 | .027*** |
| Step 2 | | | |
| Support for free speech | .173*** | .195*** | .224*** |
| Perceived effects of hate speech on others | .224*** | - | - |
| Perceived effects of violent content on others | - | .187*** | - |
| Perceived effects of sexually explicit content on others | - | - | .225*** |
| $R^2$ change | .087 | .081 | .111 |
| Total $R^2$ | .096*** | .095*** | .138*** |

*$p$ < .05, **$p$ < .01, ***$p$ < .001 ($t$ test for $\beta$, two-tailed; F test for $R^2$, two-tailed).

[a]0= Less than secondary education; 1= Secondary education or more.

[b]1= Strong Democrat, 7= Strong Republican.

[c]1= Less than 10 minutes per day, 6= More than 3 hours per day.

$\beta$ = Standardized beta from the full model (final beta controlling for all variables in the model).

For each norm-violating speech category, the regression models show significant influences of the participants' perceived effects of that category on others (PME3) on their support for using personal moderation tools to regulate that category (Model 4: hate speech – $\beta$ = .224, $p$ < .001; Model 5: violent content – $\beta$ = .187, $p$ < .001; Model 6: sexually explicit content – $\beta$ = .225, $p$ < .001), supporting H4.

## Choosing Between Platform-wide and Personal Moderation

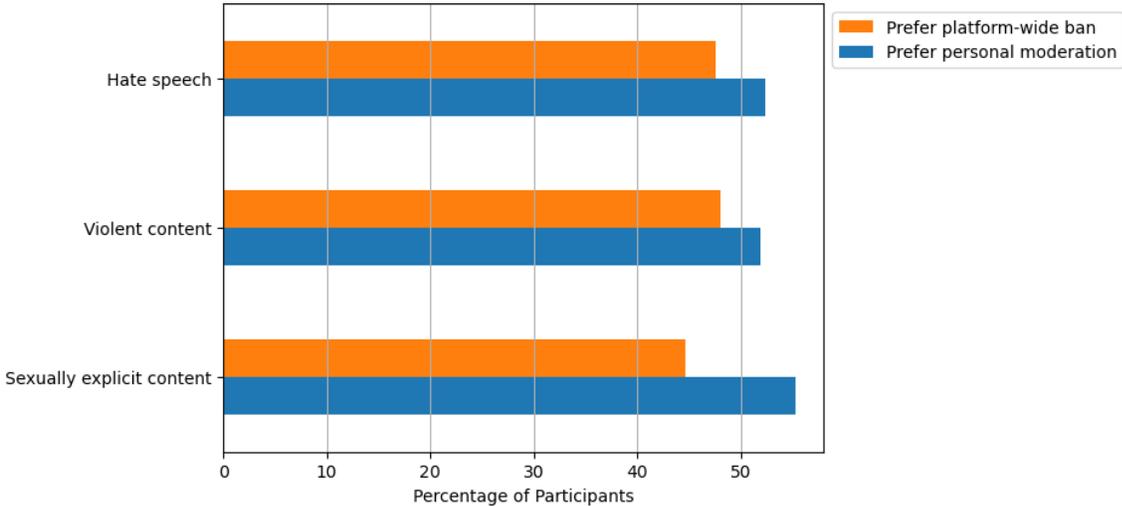

*Figure 4: Percentage of participants preferring platform-wide ban vs personal moderation to regulate hate speech, violent content, and sexually explicit content.*

Greater support for free speech has a significant positive influence on participants' support for using personal moderation tools to regulate each norm-violating category (Model 4:

hate speech – $β$ = .173, $p$ < .001; Model 5: violent content – $β$ = .195, $p$ < .001; Model 6: sexually explicit content – $β$ = .224, $p$ < .001). This answers our RQ 2.

Given a choice between platform-wide moderation and a personal moderation tool to regulate hate speech, violent content, and sexually explicit content, 52.4%, 52%, and 55.3% of participants, respectively, chose the personal moderation tool. This finding shows that more participants prefer autonomy over moderation over delegating it to platforms.

We created binomial logistic regression models to test our hypothesis 5 and answer RQ 1. For each norm-violating category, we created a model where the participants' binary choice between platform-wide and personal moderation to handle that category served as the dependent variable. In Step 1 of the three regression models, we included the control variables age, gender, education, race, political affiliation, and social media use. In Step 2, we introduced the independent variables PME3 (perceived effects on others) for that category and support for free speech. For each model, we used the Box-Tidwell procedure (Box and Tidwell, 1962; Fox, 2015) to check the assumption of linearity in the logit. We found in each case that our continuous variable *support for free speech* was not linearly related to the logit of the dependent variable. To address this, we split this variable into two ordinal categories, *high* and *low*, recoding each entry for this variable based on whether it exceeded the median value. Rerunning the Box-Tidwell procedure with this transformed *support for the free speech categorical* variable, we found that all remaining continuous independent variables in each model were linearly related to the logit of the dependent variable.

We next present these models' binomial logistic regression results (Table 5).

*Table 5: Binomial logistic regression analyses predicting support for using personal moderation tools (PMT) over platform-wide ban to regulate hate speech, violent content, and sexually explicit content (N = 984).*

| Independent Variable | Support for PMT over platform ban to regulate hate speech, Odds Ratio | Support for PMT over platform ban to regulate violent content, Odds Ratio | Support for PMT over platform ban to regulate violent content, Odds Ratio |
|---|---|---|---|
| Step 1 | | | |
| Age | .986** | .992 | .994 |
| Gender (Female) | .790 | .655** | .674** |
| Race (White) | 1.107 | 1.313 | 1.203 |
| Education[a] | .934 | .870 | .981 |
| Political affiliation[b] | 1.163*** | 1.118*** | 1.074* |
| Social media use[c] | .990 | 1.071 | 1.091* |
| Nagelkerke R² | .066*** | .066*** | .045*** |
| Step 2 | | | |
| Support for free speech[d] | 2.703*** | 2.239**** | 1.969*** |
| Perceived effects of hate speech on others | .735*** | - | - |
| Perceived effects of violent content on others | - | .752*** | - |
| Perceived effects of sexually explicit content on others | - | - | .857** |
| Total Nagelkerke R² | .165*** | .138*** | .087*** |

*$p < .05$, **$p < .01$, ***$p < .001$ ($t$ test for $β$, two-tailed; Omnibus Tests of Model Coefficients for $R^2$).

[a]0= Less than secondary education; 1= Secondary education or more.

[b]1= Strong Democrat, 7= Strong Republican.

[c]1= Less than 10 minutes per day, 6= More than 3 hours per day.

[d]0= low, 1=high.

Odds Ratio = $exp(β)$ from full model.

For each norm-violating speech category, the regression models show significant influences of the participants' perceived effects of that category on others (PME3) on their choice of using personal moderation tools over platform-enacted bans to regulate that category. Increasing PME3 was associated with a decreased likelihood of choosing personal moderation tools over platform bans (Model 7: hate speech – $exp(β)$ = .735, $p < .001$; Model 8: violent content – $exp(β)$ = .752, $p < .001$; Model 9: sexually explicit content – $exp(β)$ = .857, $p < .01$). This answers our RQ 1.

Higher support for free speech has a significant positive influence on participants' support for using personal moderation tools over platform-enacted bans to regulate each norm-violating category. Participants who showed high support for free speech have 2.703, 2.239, and 1.969 times higher odds of choosing personal moderation tools over platform-enacted bans to regulate hate speech, violent content, and sexually explicit content, respectively. Thus, H5 is supported.

# Discussion

Third-person effects (TPE) are commonly used to examine people's attitudes toward censorship and related behaviors or behavioral intentions (Davison, 1983; Gunther, 1991). This paper extends the TPE research to managing hate speech, violent content, and sexually explicit content on social media. We explored the presumed negative effects of each type of content on self (PME1) and others (PME3). The results supported the TPE hypothesis: as expected, participants perceived that hate speech, violent content, and sexually explicit content on social media exert a greater influence on others than on themselves.

We also examined how PME3 and support for free speech affected participants' consequent censorial behavior. In each case, we found that the perceived effects on others (PME3) predicted participants' support for both platform-wide and personal moderation. *This theoretically significant finding helps advance TPE research by showing that perceived effects on others play an essential role in triggering censorial behavior.* Given a choice between the platform and personal moderation, greater PME3 predicted preference for platform moderation over personal moderation in each content category. This finding indicates that when users perceive the adverse effects of a content category on the public, they desire platforms to take site-wide actions on that content rather than regulate it for themselves. This suggests a public appetite for platforms to take on the responsibility to protect vulnerable others from content deemed to be egregious, even at the expense of personal moderation control.

We found only partial evidence for the relationship between free speech and platform-wide moderation support. While the connection is significant and negative for hate speech, it is not significant for violent and sexually explicit content. This result is consistent with the mixed findings for free speech support as a negative predictor of supportive attitudes towards platform censorship observed in prior literature (Guo and Johnson, 2020). On the other hand, *we found that support for free speech predicted support for the use of personal moderation for regulating each inappropriate speech category.* This suggests that people may perceive personal moderation tools not as an infringement on the free speech of others but simply as according them greater personal agency to shape what they see. Further bolstering this interpretation is our finding that given a choice between platform and personal moderation, support for free speech predicts support for personal moderation in each content category.

Other significant effects less central to the hypotheses being tested were also found. We found that age was positively related to support for platform moderation of hate speech and sexually explicit content, but not violent content. Females supported platform moderation of each speech category more than males. Democrats were more likely than Republicans to support platform moderation of hate speech and violent content, but not sexually explicit content. Regarding support for personal moderation, race, political affiliation, and social media use were significant predictors for the sexually explicit content category; however, no control variables significantly predicted personal moderation of hate speech or violent content. Future research could offer more insights into these results by examining how users' political ideology, media consumption, parental concerns, and prior experiences on social media shape their moderation preferences.

The evidence presented here has important implications for how platforms govern their sites. We show that the public supports both platform and personal moderation of norm-violating categories such as pornography and war violence in part because it overestimates these categories' effects on others. Therefore, company-wide moderation decisions and public debates concerning free speech and its limitations must recognize and account for third-person effects. It also points to an urgent need to measure *actual* social media effects as opposed to the *perceived* effects of different content types. Even during our separate interview studies on online harms,[iv] we noticed that participants tend to advocate for specific moderation initiatives based on their perceptions of what *others* might need. To account for biases due to third-person effects, scholars must examine online content from the perspectives of users themselves – a topic on which they are an expert – rather than on abstract others whose actual needs may considerably differ.

We found that given a choice between platform-wide and personal moderation, participants prefer the latter to regulate each norm-violating speech category (Figure 4). However, many platforms currently do not support or poorly support personal moderation (Jhaver et al., 2022). Our results support a scheme where both platform and personal moderation are available and robust. Platforms could reduce their site-wide bans of borderline content (e.g., content that does not generate high PME3) and instead empower users with personal moderation tools that allow specifying their moderation preferences for that content.

Several limitations of this study should be recognized. The survey design prohibited us from exploring in depth the motivations for specific perceptions. Furthermore, we cannot make conclusive statements about causal relations in a cross-sectional study. We asked participants to respond to questions about speech categories that could be broadly interpreted; we chose this instead of presenting specific instances of each speech category to increase the generalizability of our findings. Still, different users may have different perceptions of what counts as hate speech, violent content, or sexually explicit content. Prior moderation research has recognized this as a complex challenge in social media regulation (Jhaver et al., 2018). We provided definitions of each speech category in the survey to clarify the scope of each category to our participants.

Nevertheless, further research on user perceptions of stimulus-based designs that present preselected instances of each norm-violating speech category to participants would provide valuable insights. We presuppose a setup where the platform defines inappropriate speech categories and labels what posts fall under them. However, each site must examine whether its moderation implementation could accommodate this arrangement. Finally, we did not ground our survey questions in a specific platform to increase the generalizability of our results. Studies focused on particular social media sites can uncover whether attitudes towards specific platforms influence users' perceptions of moderation actions.

---

[i] We will specify the Institute name after the peer review process is completed.
[ii] https://lucidtheorem.com
[iii] The First Amendment to the United States Constitution states: "*Congress shall make no law respecting an establishment of religion, or prohibiting the free exercise thereof; or abridging the freedom of speech, or of the*

*press; or the right of the people peaceably to assemble, and to petition the Government for a redress of grievances."*

[iv] We will add citations after the peer review.